\documentclass[aps,twocolumn,floatfix,showpacs,superscriptaddress]{revtex4-1}
\usepackage{graphicx}
\usepackage{natbib}
\usepackage{epstopdf}
\usepackage{amsmath}
\usepackage{amssymb}
\usepackage{mathbbol}
\usepackage{braket}
\usepackage{hyperref}

\usepackage{xcolor}

\newcommand{\beq}{\begin{equation}}
\newcommand{\eeq}{\end{equation} \smallskip}
\newcommand{\beqy}{\begin{eqnarray}}
\newcommand{\eeqy}{\end{eqnarray} \smallskip}

\newcommand{\expect}[1]{\langle #1 \rangle}
\newcommand{\lsim}{\lower.7ex\hbox{$\;\stackrel{\textstyle<}{\sim}\;$}}

\begin{document}

%% title, authors etc.
\title{Simultaneous bistability of qubit and resonator in circuit
  quantum electrodynamics}

\author{Th. K. Mavrogordatos}\email{t.mavrogordatos@ucl.ac.uk}
\affiliation{Department of Physics and Astronomy, University College London,
Gower Street, London, WC1E 6BT, United Kingdom}
\author{G. Tancredi}
\affiliation{Clarendon Laboratory, University of Oxford, Parks Road,
  Oxford, OX1 3PU, United Kingdom} 
\author{M. Elliott}
\affiliation{Advanced Technology Institute and Department of Physics,
  University of Surrey, Guildford, GU2 7XH, United Kingdom} 
\author{M. J. Peterer}
\affiliation{Clarendon Laboratory, University of Oxford, Parks Road,
  Oxford, OX1 3PU, United Kingdom} 
\author{A. Patterson}
\affiliation{Clarendon Laboratory, University of Oxford, Parks Road,
  Oxford, OX1 3PU, United Kingdom} 
\author{J. Rahamim}
\affiliation{Clarendon Laboratory, University of Oxford, Parks Road,
  Oxford, OX1 3PU, United Kingdom} 
\author{P. J. Leek}
\affiliation{Clarendon Laboratory, University of Oxford, Parks Road,
  Oxford, OX1 3PU, United Kingdom} 
\author{E. Ginossar}
\affiliation{Advanced Technology Institute and Department of Physics,
  University of Surrey, Guildford, GU2 7XH, United Kingdom} 
\author{M. H. Szyma\'nska}
\affiliation{Department of Physics and Astronomy, University College London,
Gower Street, London, WC1E 6BT, United Kingdom}

\date{\today}


\begin{abstract}
  We explore the joint activated dynamics exhibited by two quantum degrees of freedom: a cavity
mode oscillator which is strongly coupled to a superconducting qubit in the strongly coherently
driven dispersive regime. Dynamical simulations and complementary measurements show a range of parameters where both the cavity and the qubit exhibit sudden simultaneous switching between two metastable states. This manifests in ensemble averaged amplitudes of both the cavity and qubit exhibiting a partial coherent cancellation. Transmission measurements of driven microwave cavities coupled to transmon qubits show detailed features which agree with the theory in the regime of simultaneous switching.
\end{abstract}

\maketitle

\section{Introduction}The generalized Jaynes-Cummings (GJC) model provides a simple basis for
describing the interactions between a quantized electromagnetic field
and multilevel atoms. Its nonlinearity lies
at the heart of cavity quantum electrodynamics (cavity QED), where natural atoms are coupled to cavity photons
\cite{cavQED}, and circuit quantum electrodynamics (circuit QED), where artificial atoms are coupled to resonators of various dimensionalities \cite{TransmonPaper, cQED1, cQED2, JJ, WallsBook}. The Jaynes-Cummings (JC) interaction also emerges in areas of current interest such as optomechanics in the
linearized regime \cite{ReviewOptomechanics} and in the Bose-Hubbard
model \cite{PhotonBlockade}. There is a large body of work on the
resonant and strong-coupling regime of the driven-dissipative JC oscillator \cite{NatomsBist, Remnants},
where driving induces a dynamical Rabi splitting \cite{QuantumTrajectory, BishopRabi}. The
high excitation strong-dispersive regime is also of great interest, for example, in the context of amplifiers \cite{JBA}, squeezing associated with the parametric oscillator
\cite{CarmichaelBook2} and the implementation of qubit readout schemes \cite{autoresonance, Boissonneault, BishopJC, Reed}. In this context, fluctuation-induced switching between metastable states in the driven-dissipative GJC system, which involves two quantum degrees of freedom, has not been directly studied but the theory of quantum activation motivates the interest in this scenario \cite{DykmanJETP, DykmanBook, Peano, Kamenevbook, Maier, DykmanPLA, Graham}. 

\begin{figure}[!ht]
\centering
\includegraphics[width=3.1in]{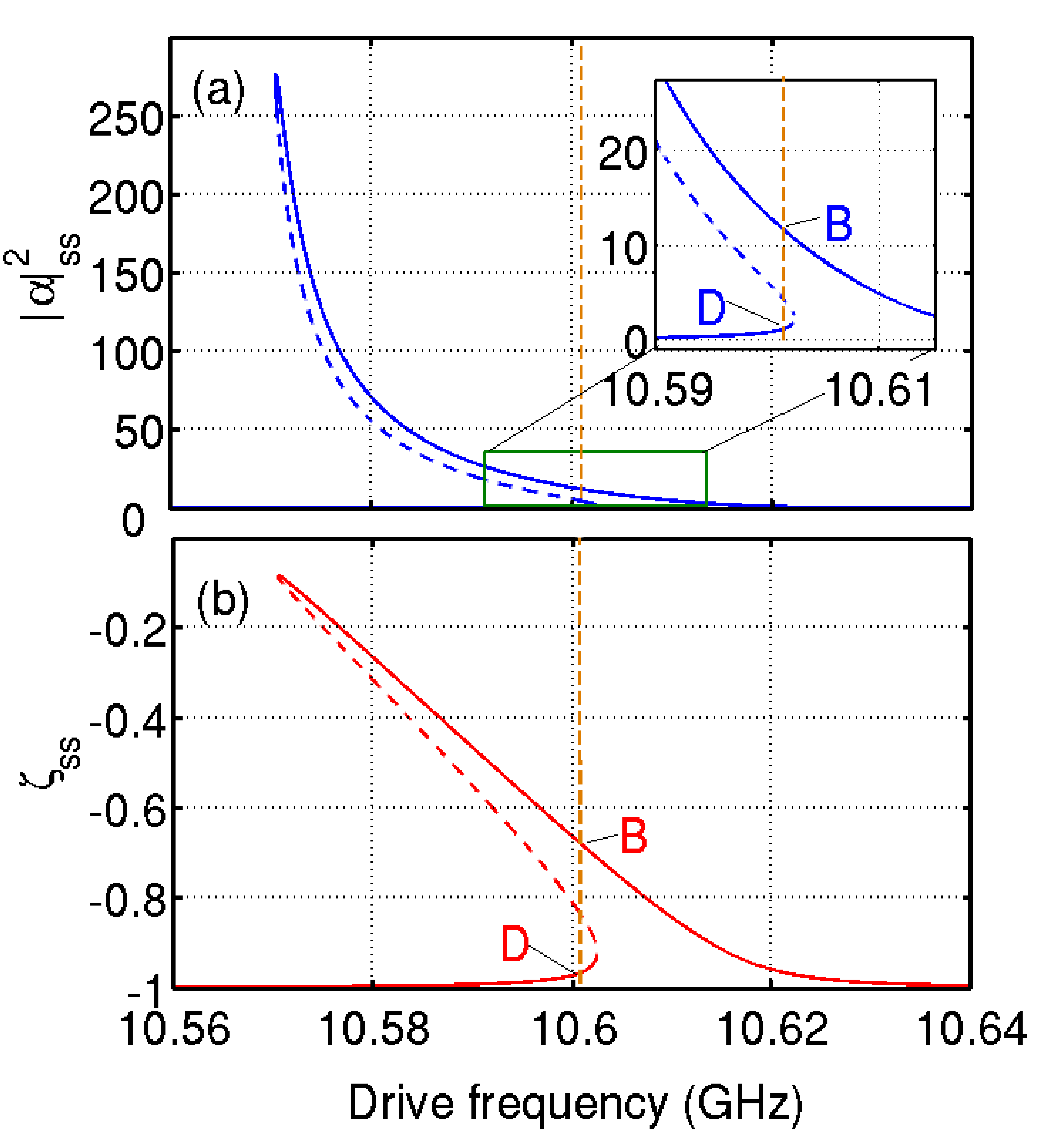}
\caption{Maxwell-Bloch steady-state bistability of the JC oscillator in the dispersive regime \cite{WallsBook} for the coupling strength to detuning ratio $g/\delta=0.14$, the drive amplitude to photon loss ratio $\varepsilon_d/(2\kappa)=25/3$ and the photon loss to spontaneous emission ratio  $2\kappa/\gamma=12$. Here, ${\rm B (D)}$ denote the bright (dim) semiclassical states. \textbf{(a)} Intracavity field amplitude in the steady state (with the region of interest zoomed in the inset). \textbf{(b)} The corresponding atomic inversion $\zeta=\braket{\sigma_z}$. The broken orange line indicates the driving frequency at $10.6005\,$GHz, and the dashed lines in each curve depict the unstable branch.}
\label{fig:MF}
\end{figure}

\begin{figure}[!ht]
\centering
\includegraphics[width=3.5in]{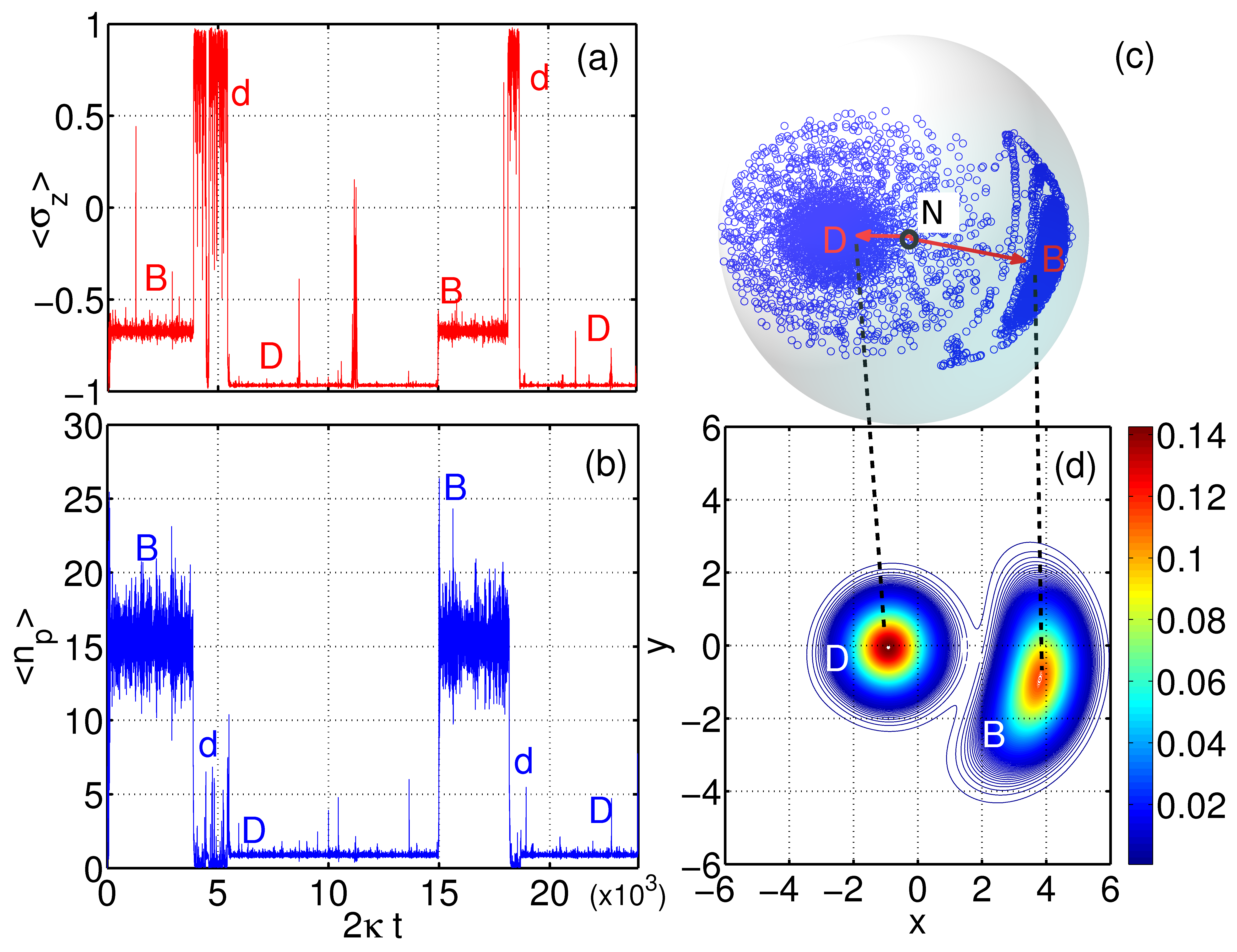}
\caption{Quantum-activated simultaneous cavity and qubit switching illustrated using the JC model in the dispersive regime for $g/\delta=0.14$, $3 \varepsilon_d/(2\kappa)=25$, $2\kappa/\gamma=12$ and $N_{\textrm{crit}} \simeq 13$. Here, ${\rm B (D)}$ denote the bright (dim) metastable states and ${\rm d}$ denotes the nonclassical dark state. \textbf{(a)} Atomic population inversion $\Braket{\sigma_{z}}$. \textbf{(b)} The accompanying intracavity photons $\Braket{n_p}=\Braket{a^{\dagger}a}$ as a function of the dimensionless time $2\kappa t$ for a single quantum trajectory. The trajectories in (a) and (b) depict simultaneous switching between the bright (B) and dim (D) states. \textbf{(c)} Illustration depicting the two metastable state distributions in the Bloch sphere (as viewed from the north pole indicated by the letter N). Data points corresponding to the dark state are omitted for clarity. The red arrows point to the two metastable states (B and D). \textbf{(d)} Contour plot of the joint quasidistribution function $Q(x + iy)$ for $f_d=\omega_d/2\pi=10.6005\,$ GHz, as indicated in Fig. \ref{fig:MF}, showing two peaks corresponding to two semicoherent states, indicating the presence of cavity bimodality.}
\label{fig:MEU} 
\end{figure}

In this paper, we demonstrate that in a nonlinear intermediate driving regime of circuit QED, the system dynamics exhibits a simultaneous
bistability of the qubit and resonator. We support this claim both theoretically, with analytical and numerical results, as well as with experimental measurements obtained from a circuit QED device, consisting of a transmon qubit coupled to a 3D microwave cavity \cite{Paik}. In particular we report the following findings.
\textbf{(1)} The switching process occurs simultaneously for the two coupled quantum oscillators. \textbf{(2)} The ensemble averaged amplitudes of both cavity and qubit exhibit a coherent partial cancellation. Such cancellation, predicted theoretically for a single nonlinear mode by Drummond and Walls in \cite{DrummondWallsKerr}, is here verified experimentally and shown to occur for both coupled oscillators.  
\textbf{(3)} The Duffing oscillator model with one quantum degree of freedom is not sufficient to account for the observed cavity nonlinearity when the two coupled quantum degrees of freedom are involved in the switching process.
We also note that, the JC bistable \emph{semiclassical} intracavity amplitude $|\alpha|_{\rm ss}^2$ (with $\alpha=\braket{a}$) in the steady state, plotted 
in Fig. \ref{fig:MF}(a), does not show any coherent cancellation feature, nor does the bistable average atomic inversion $\zeta_{\rm ss}=\braket{\sigma_z}_{\rm ss}$ shown in Fig. \ref{fig:MF}(b). Instead, both curves are skewed Lorentzians with the position of their peaks approaching the bare cavity frequency for increasing drive. By increasing the drive strength beyond what is shown in Fig. \ref{fig:MF}(a) one finds a critical point in the phase space between bistable and linear behavior, lying on the line where the frequency of the drive equals the bare cavity resonance frequency. Beyond that point, the system behaves as a linear oscillator, in contrast to the corresponding Duffing oscillator \cite{BishopJC}, and exhibits no bistability.

\section{Theoretical models} In order to develop a comprehensive
understanding of the system response, we will consider several
different theoretical models: (i) multilevel transmon-cavity GJC
model --- the most complete in the context of superconducting devices;
(ii) two-level `atom'-cavity JC model which is universal to many
strong light-matter coupling scenarios, and (iii) a simplified
dressed-cavity Duffing oscillator approximation. To complement the
numerical simulations of the three models, we additionally compare the
results for the cavity transmission with those coming from an
analytical formula, obtained by modeling the transmon itself as a
nonlinear oscillator.

We will now define the system Hamiltonians associated with models (i)-(iii).
When one cavity field mode of frequency $\omega_c$ (with corresponding photon annihilation and creation operators $a$ and $a^{\dagger}$ respectively) is coupled to a multilevel system with unperturbed states $\Ket{n}$, the
coherently-driven GJC Hamiltonian can be written as (setting $\hbar=1$)
\cite{TransmonPaper}
\begin{multline}\label{TransmonJC}
H_{\rm{GJC}}^{\text{(i)}}= \omega_c a^{\dagger}a +  \sum_{n} \omega_{n} \Ket{n}
\Bra{n} +  \sum_{m,n} g_{mn} \Ket{m}\Bra{n} (a + a^{\dagger})  \\ 
+ i  \varepsilon_d (a^{\dagger}e^{-i\omega_d t} - ae^{i \omega_d t}), 
\end{multline}
where $\varepsilon_d$ is the strength of a monochromatic external
field with frequency $\omega_d$ driving the cavity mode. The sum in
the third term describes the interaction and is customarily modified
to $ \sum_m  g_{m, m+1} \left( \Ket{m} \Bra{m +1}a^{\dagger} +
  \textrm{h.c.}\right)$ in the rotating wave approximation (RWA). The
interaction energies in the RWA have the approximate form $ g_{mn}
\approx g\sqrt{m+1}\,\,\delta_{m+1,n}$, with $g$ being the dipole coupling strength.
Depending on the range of $n$ and
the form of $\omega_n$ we can distinguish the two-level atom ($n=1$,
$m=0$, $\omega_n=\omega_q$) -- the JC model -- from a transmon ($n=1,2,
\ldots ,N_{\textrm{max}} $, $\omega_n=\omega(n)$) -- the GJC model. The JC model in the RWA reads
\begin{align}\label{aJC}
H_{\rm{JC}}^{\text{(ii)}}=\omega_{c} a^{\dagger}a + \frac{1}{2}\omega_{q}
\sigma_{z} + g (a^{\dagger} \sigma_{-} + a \sigma_{+}), 
\end{align}
with $\sigma_{\pm}$ the raising (lowering) pseudospin operators and
$\sigma_z=2\sigma_{+} \sigma_{-} -1$ the inversion
operator. In the presence of dissipation, the cavity mode
is damped at a rate $2\kappa$ \cite{doublerate} while spontaneous emission is present at
a rate $\gamma$ for a qubit dephased at a rate $\gamma_{\phi}$.
\begin{figure}[!ht]
\includegraphics[width=3.0in]{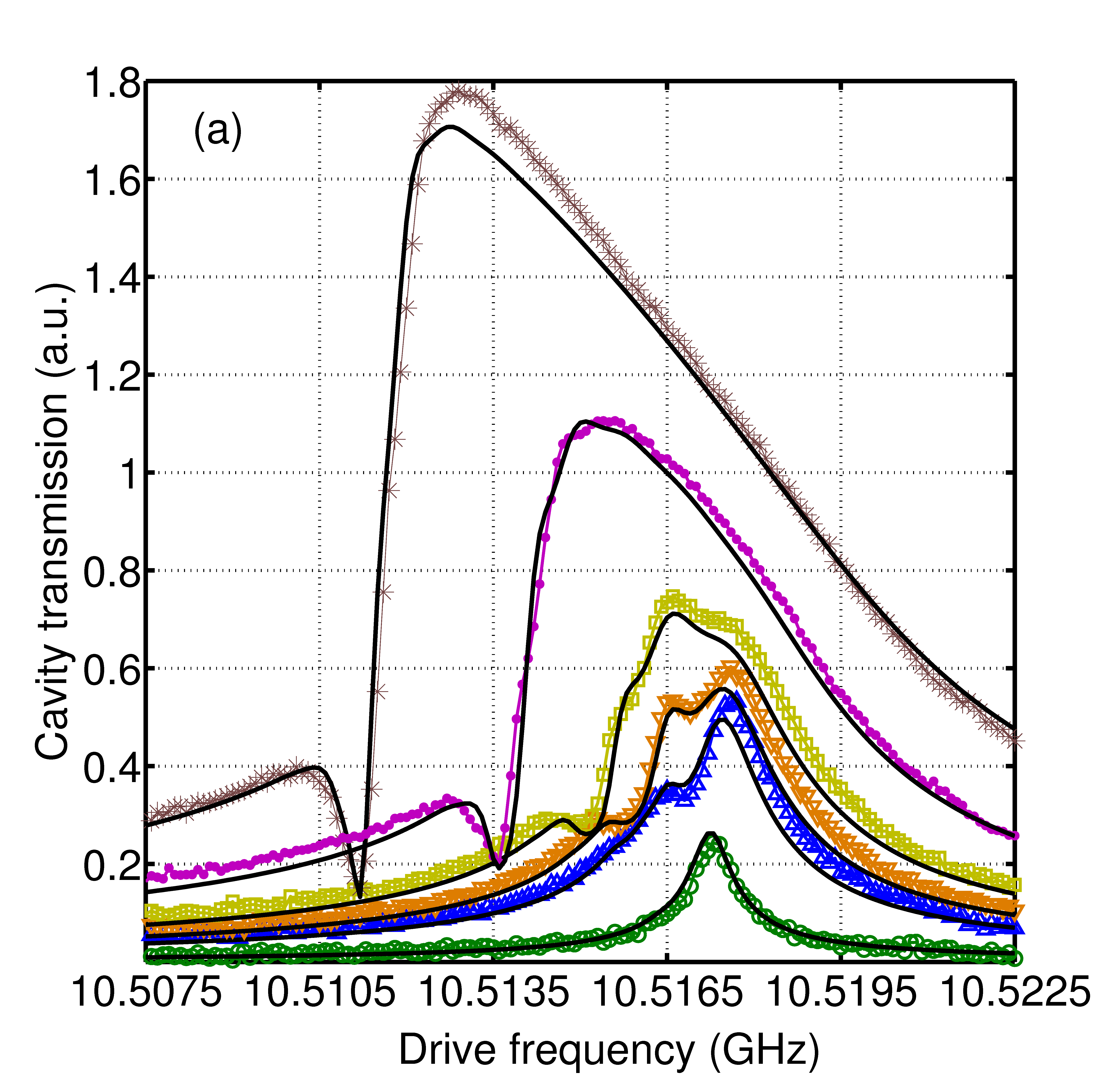}
\includegraphics[width=3.0in]{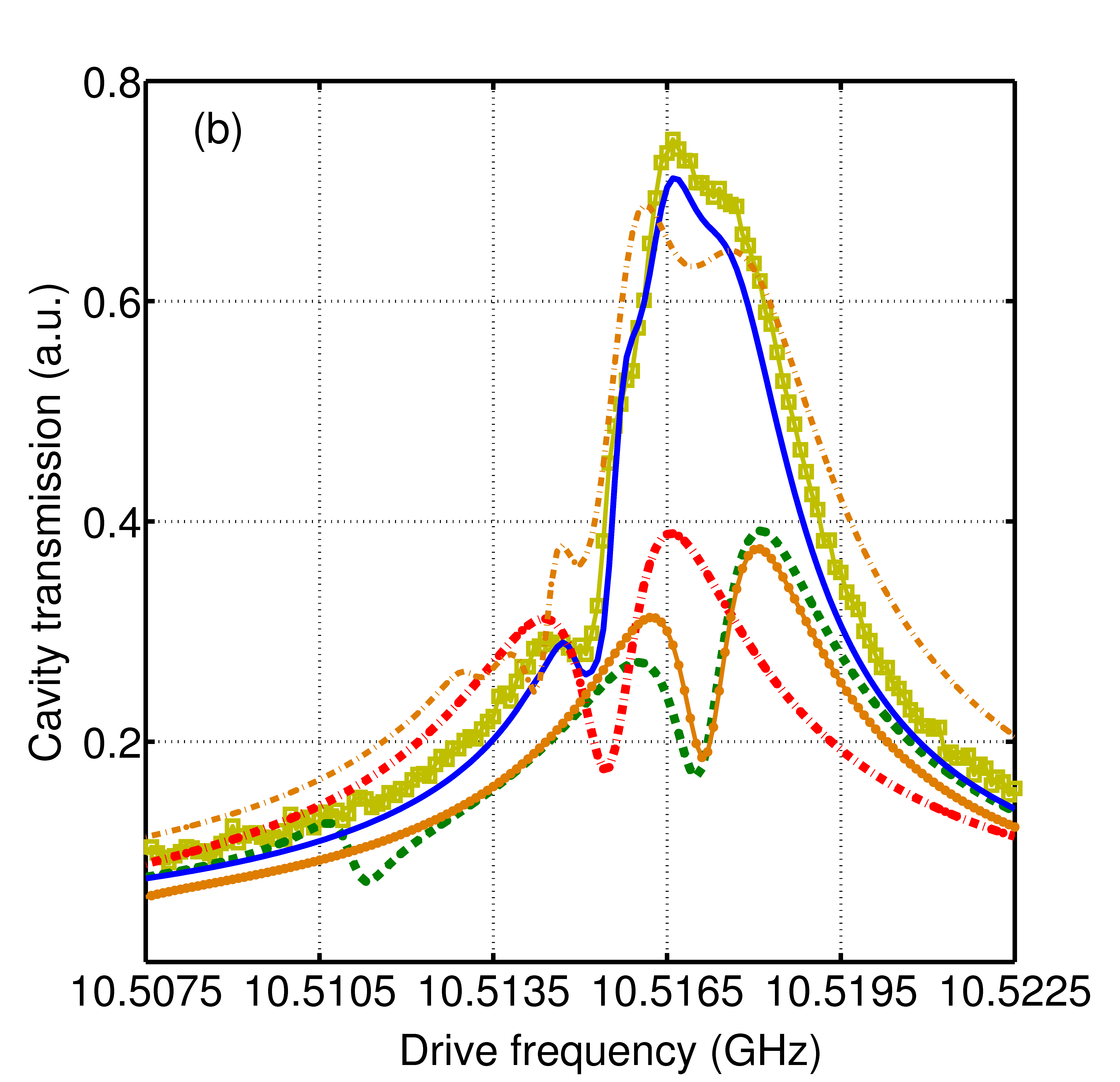}
\caption{Theory and experiment showing the
  cavity nonlinearity for increasing drive strength. \textbf{(a)}
  The experimentally measured transmission amplitude (in patterned lines) of the device $D_1$, consisting of a cavity resonator with bare frequency
  $f_c=\omega_c/2\pi=10.426\,$GHz coupled to a transmon qubit with lower transition frequency $f_q=\omega_q/2\pi=9.442\,$GHz, for six increasing values of the driving strength ($-66\,$dBm: green $\circ$, $-52\,$dBm: blue $\triangle$, $-49\,$dBm: orange $\triangledown$, $-46\,$dBm: yellow $\square$, $-41\,$dBm: pink $\bullet$ and $-36\,$dBm: brown $\ast$) superimposed on top of the theoretical predictions of the GJC model with four levels (shown in solid black lines).
  \textbf{(b)} Cavity transmission as a function of the driving frequency for all the considered theoretical models: GJC in solid \textbf{blue}, JC in dashed \textbf{green}, its Duffing reduction in dot-dashed \textbf{red} and the effective Fokker-Planck model in \textbf{orange} with full circles for the experimental transmon anharmonicity coefficient $\chi/2\pi=-150\,$MHz, and a dot-dashed line for $\chi/2\pi=-20\,$MHz, alongside the experimental data in yellow $\square$ for the driving power of $-46\,$dBm. It is clear that (only) the GJC model shows excellent agreement with the experimental data.}
\label{fig:GJCcomp}
\end{figure}
It is possible to approximate the Hamiltonian of Eq. \eqref{aJC}
further in the strongly dispersive regime, defined by the 
strong detuning $\delta=\left|\omega_c - \omega_q\right| \gg g$
between the two coupled oscillators. 
Under an appropriate decoupling transformation
\cite{DecouplingT, Peano} $H_{\rm JC}$ can be
recast in the form $H'_{\rm JC}= \omega_c a^{\dagger}a + (1/2)
(\omega_c - \Delta) \sigma_z$, involving the operator
$\Delta=\sqrt{\delta^2 + 4g^2 \mathcal{N}}$ where
$\mathcal{N}=a^{\dagger}a + \sigma_{+} \sigma_{-}$ is the operator of
the total number of excitations $N$. In the dispersive regime,
provided that $N/N_{\rm crit} \ll 1$  \cite{ComparisonRD}, we can expand up to the
quartic order in the field variables. After normal
ordering we obtain the following dressed-cavity Duffing oscillator Hamiltonian
\begin{equation}\label{Duffingapprox}
H_{\text{D}}^{\text{(iii)}}= \left(\omega_c + \frac{g^4}{\delta^3} -
  \frac{g^2}{\delta}\sigma_z + 2 \frac{g^4}{\delta^3}\sigma_z
\right)a^{\dagger}a
+  \frac{g^4}{\delta^3}\sigma_z{a^{\dagger}}^2 a^2,
\end{equation}
where setting $\sigma_z=-1$ is a justifiable approximation for
low enough driving amplitudes, yielding a bistable quantum Duffing
oscillator \cite{driveterms}. The third term in the parentheses is the leading-order term $ 
\chi^{(0)}_c=(g^2/\delta) \sigma_z$ which is the familiar
Stark shift \cite{Blais2004} and provides a valuable tool for qubit
readout (with $\left|\chi^{(0)}_c\right| \gg \kappa$). Here, in the bifurcating dispersive region we are studying, the following hierarchy of scales applies
\cite{BishopJC}: $\gamma, \gamma_{\phi} \ll 2\kappa \ll \varepsilon_d
\lsim g^2/\delta \ll g < \delta \ll \omega_c$.  The intracavity
excitation number is of the order of $N_{\rm crit}$, where this
perturbation expansion is not strictly valid. However, as we will see
later, it gives qualitatively meaningful results.

Having defined the different model Hamiltonians under consideration,
we now evolve the corresponding master equations (MEs) in the finite
Hilbert state basis numerically, starting from a Fock state of zero photons and
the qubit in the ground state, until it reaches a steady state.  Note that the steady state
obtained is independent of the choice of the initial conditions.

\section{Activated dynamics in the dispersive regime} Driving the
system beyond the low power regime has a profound effect on the
response. We illustrate this fact in Fig. \ref{fig:MEU}, where we depict the
qubit inversion $\braket{\sigma_z}$ in Fig. \ref{fig:MEU}(a),
the photon cavity number $\braket{n_p}=\braket{a^{\dagger}a}$ in Fig. \ref{fig:MEU}(b), alongside their associated cavity
quasidistribution function in Fig. \ref{fig:MEU}(d), employing the exact ME simulations
[Fig. \ref{fig:MEU} (d)] and single quantum trajectories from the
stochastic Schr{\"o}dinger equations (SSE) using the second-order weak
scheme in the diffusive approximation \cite{BreuerBook, PlatenBook}
[Figs. \ref{fig:MEU}(a) and \ref{fig:MEU}(b)]. Single quantum trajectories, corresponding to
the unravelling of the ME for the JC Hamiltonian, depict the switching
between the two metastable semiclassical states as a result of
quantum fluctuations. The switching occurs simultaneously for the
qubit [Fig. \ref{fig:MEU}(a)] and the cavity [Fig. \ref{fig:MEU}(b)]. The
corresponding photon histogram shows quasi-Poissonian statistics
obeyed by the two metastable states, one with mean photon
occupation of the order of $N_{\textrm{crit}}$ (called `bright'
state), one with mean occupation of about a photon (called `dim'
state) as well as the distribution of a nonclassical (called `dark')
state (for more details see the Supplementary Information). In Fig. \ref{fig:MEU}(c)
we draw a sketch illustrating the qubit distribution in the steady
state, as viewed from the north pole of the Bloch sphere, for a single quantum trajectory. Red arrows point to the two
semiclassical qubit states, corresponding to the
two metastable quasicoherent cavity states [depicted by color
contour plots in Fig. \ref{fig:MEU}(d)] between which quantum-activated switching takes
place. The $Q$ function plot in Fig \ref{fig:MEU}(d) also shows the position in
the phase space of the coherently cancelling states. The equal height of the
$Q$ function peaks indicates the boundary of a first-order dissipative
quantum phase transition \cite{PhotonBlockade, Protocol}. This
transition is marked by the switching rates to the bright and to the
dim state being of the same order of magnitude \cite{PhotonBlockade,
  DykmanJETP}. In our case, as well as in the exact photon statistics
section of \cite{DrummondWallsKerr}, switching is induced by quantum
fluctuations only, as the thermal bath to which the system is coupled
is at zero temperature. The trajectories depicted for illustration in Fig. \ref{fig:MEU} evidence sudden simultaneous jumps. Bistability and synchronisation were studied for the two-level Rabi model in \cite{Zhirov}. Note that, in contrast to the cavity field and
$\left|\Braket{\sigma_{-}}\right|$, the exact ME results for the photon number and
$\Braket{\sigma_{z}}$ show no coherent cancellation in the steady-state response. The mean-field behavior, depicted in Fig. \ref{fig:MF}, shows further that the coherent cancellation is purely a quantum effect at zero temperature, occurring when forming the ensemble averaged quantities, and is already present in the most approximate dressed-cavity Duffing model, even if the qubit is unmonitored. 

\section{The nonlinear resonator transmission lineshape} In
Fig. \ref{fig:GJCcomp}(a) we compare theoretical and experimental
transmission amplitudes of a 3D cavity with embedded transmon (device
$D_1$, details in the Supplementary Information) for different driving
strengths. We observe that, as the driving power is increased, the
experimental cavity lineshape develops nonlinear features and a
coherent cancellation dip appears. We find perfect agreement with the
GJC model.  In Fig. \ref{fig:GJCcomp}(b) we show the cavity transmission
for the intermediate drive power of $-46\,$ dBm for all the models
discussed. We observe that the JC model predicts the split of the
main peak at the correct position, as opposed to its Duffing
reduction, yet fails to capture the position of the dip emerging at a
lower frequency. The GJC model with four transmon levels can resolve
all the details necessary for a quantitative comparison and provides indeed
the most complete description of the cavity nonlinearity.

The behavior of the GJC oscillator depends strongly on the drive strength and frequency, and their relation to the coupling and dissipation rates \cite{CarmichaelBook2, BishopJC}. We find theoretically that the coherent cancellation dip in transmission, discovered by Drummond and Walls for the Duffing oscillator \cite{DrummondWallsKerr}, appears in the dispersive response of both the cavity and the qubit within the full nonlinearity of the JC model, where the departure from the mean-field predictions is appreciable and the Duffing oscillator approximation is no longer valid. For the Duffing oscillator the dip is present in the first moment of the field operator, $\left|\braket{a}\right|$, calculated using the generalized $P$ representation \cite{DrummondWallsKerr}. It is purely a phase effect as the dip does not appear in the number of intracavity photons in the steady state $\Braket{n_{\textrm{p}}}=\Braket{a^{\dagger}a}$. The coherent cancellation dip appears as well in the qubit projection $\left|\braket{\sigma_{-}}\right|$ (see the Supplementary Information for more details). The presence of this dip in the cavity response has also been observed in our experimental measurements, which depict the development of nonlinearity for increasing drive strengths within the region of bistability. Similar
cancellation effects appear also in classical dissipative systems out of equilibrium, in the presence of thermal fluctuations \cite{DrummondWallsKerr}. The observed dip, appearing progressively in a measurement of {\em complex amplitude}, is due to the phase differences between the two metastable states. In the experimental response we can also discern a split of the main peak, alluding to dynamical Rabi splitting \cite{BishopRabi}. The position of the dip shifts to the lower frequencies with increasing drive strength, while the split gradually fades away in favour of a Duffing-type profile.

In order to gain a further insight, we undertake an analytical approach by
identifying an effective Hamiltonian to produce a (second-order) Fokker-Planck equation (FPE) for the transmon, following the adiabatic elimination of the
cavity. FPEs in the generalized $P$ representation can be used to solve
exactly for the steady state of quantum systems subject to the
so-called `potential conditions', and have been used to study single
nonlinear resonator systems \cite{WallsBook}. For our two-oscillator
model, these conditions are not satisfied, yet in the limit $2\kappa
\gg \gamma, \gamma_{\phi}$ the cavity can be eliminated in a similar
fashion to the method of \cite{DrummondWallsHG}. This process leaves a
FPE for an effective one-oscillator system, which resembles a driven,
damped quantum Duffing oscillator with anharmonicity $\chi$ \cite{DrummondWallsKerr} but with
parameters that are nontrivial functions of those of the full
system. Full details of this method can be found in
\cite{ElliottElimination}. The first moment of the cavity field in the steady state is
\begin{equation}\label{momentD}
  \expect{a} = \frac{2}{\tilde{\gamma}_c}\left[\varepsilon_d
    -\frac{\tilde{\varepsilon}g}{\chi c}\frac{~_0{F}{_2}
      \left(c+1,c^*;2\left|\displaystyle\frac{\tilde{\varepsilon}}{\chi}\right|^2\right)}{~_0{F}{_2}\left(c,c^*;2\left|\displaystyle\frac{\tilde{\varepsilon}}{\chi}\right|^2\right)}\right],      
\end{equation}
where $~_0{F}{_2}(x,y;z)$ is a generalized hypergeometric function,
and we have defined effective decay constants for the cavity
$\tilde{\gamma}_c= \kappa + 2i\Delta\omega_c$ and transmon
$\tilde{\gamma}_q = \gamma +2i\Delta\omega_q + 2g^2/\tilde{\gamma}_c $
respectively (with $\Delta\omega_{c(q)}=\omega_{c(q)} - \omega_d$),
effective drive strength $\tilde{\varepsilon} =
-2ig\varepsilon_d/\tilde{\gamma}_c $ and also
$c=\tilde{\gamma}_q/(2i\chi)$. The calculated transmission amplitude via Eq. \ref{momentD} is plotted in Fig. \ref{fig:GJCcomp}(b) and
compared to the exact ME results alongside the experimental data. The effective Fokker-Planck model exaggerates the actual nonlinearity in this regime, yet a lower value of $\chi$ allows us to capture the essential features of the full transmon-cavity-driven interaction (more details in the Supplementary Information).

\section{Discussion and concluding remarks} We have examined the dispersive
interaction of a single qubit and a microwave cavity mode, tracking
nonlinearity with increasing drive power. When the regime of
bistability is reached, simultaneous switching events allow for both
of the metastable states to participate even at zero temperature. 
Their different phases cause the `dip' in coherent transmission, for which we have presented theoretical and experimental evidence. Interestingly, the dim quasicoherent state is preceded by a lower amplitude nonclassical state which is not predicted by the mean-field treatment. This state is characterized by very low photon numbers and intense
fluctuations in the qubit inversion, which occupies now the north pole of the Bloch sphere. For high excitations, beyond the
Duffing oscillator regime, both the cavity and the qubit participate in the switching, and the quantitative comparison
with the experiment necessitates the inclusion of more than two levels of the transmon. The superconducting devices we have considered
serve as examples of quantum activation with more than one quantum oscillator.

The data underlying this work is available without restriction \cite{SurreyDep}. 

\vspace{3mm}

\textbf{Acknowledgements} Th.~K.~M. wishes to thank H.~J.~Carmichael and L.~Riches for
  instructive discussions. Th.~K.~M. and M.~H.~S. acknowledge
  support from the Engineering and Physical Sciences Research Council
  (EPSRC) under grants EP/I028900/2 and
  EP/K003623/2. E.~G. acknowledges support from the EPSRC under grant
  EP/L026082/1. P.~J.~L. acknowledges support from the EPSRC under grants EP/J001821/1 and EP/M013243/1.  

\vspace{3mm}

\centerline{$\star \star \star \star \star$}

\vspace{5mm}

\centerline{\textbf{SUPPLEMENTARY INFORMATION}}

\vspace{2mm}

\section{Experimental setup}

\begin{figure*}[h]
\centering
\includegraphics[width=5.8in]{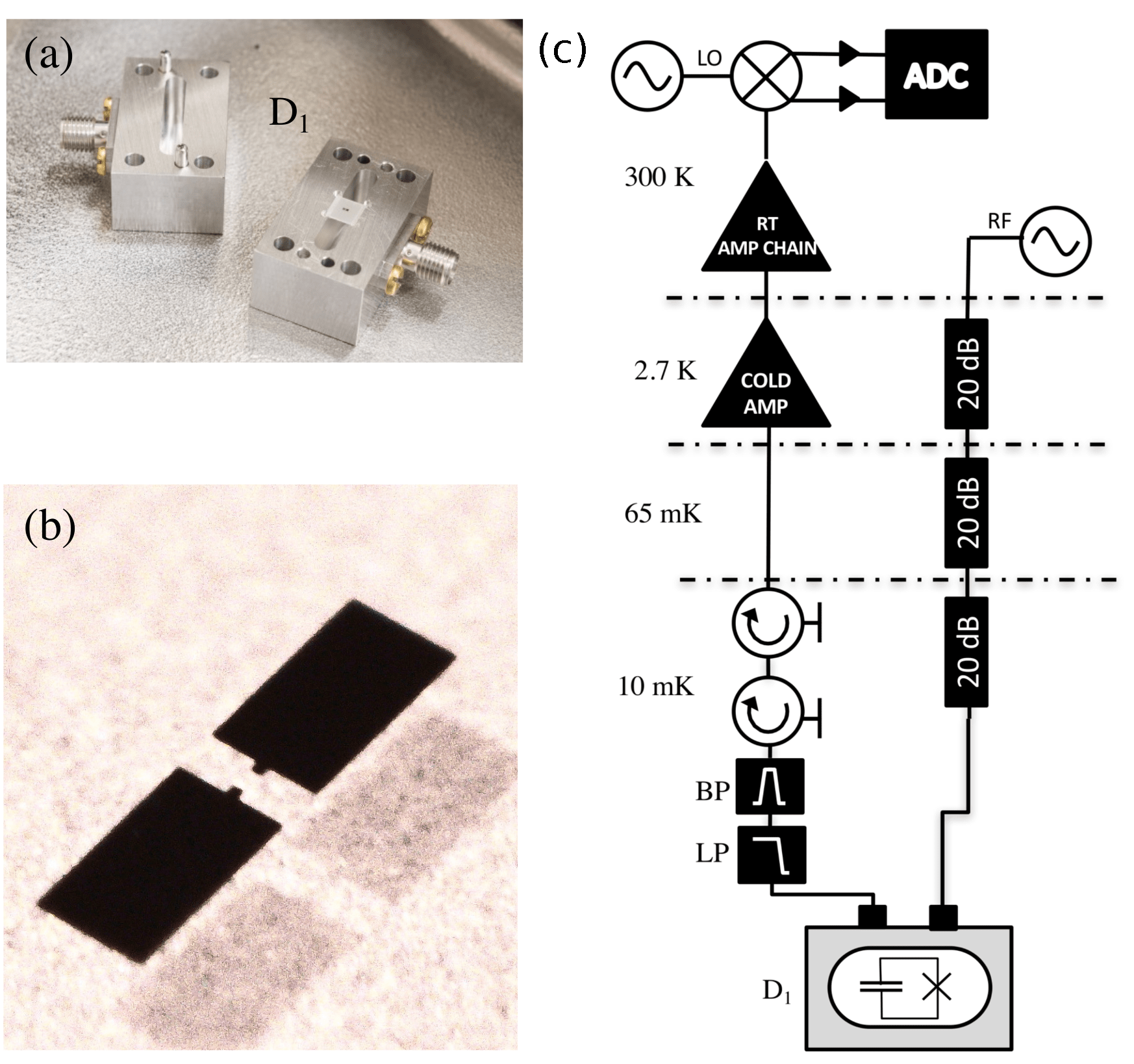}
\caption{ \textbf{(a)} The cross-section of device $D_{1}$: a
  two-ports Al microwave cavity, which holds a lithographically
  patterned Al transmon qubit on a sapphire substrate. The cavity is
  placed inside a magnetic shield at base temperature. \textbf{(b)} A
  picture of the transmon qubit consisting of two Al \textbf{(a)} pads
  of dimensions $350\,\mu$m by $450\,\mu$m connected by an Al/AlOx/Al
  Josephson junction. \textbf{(c)} The experimental setups used to
  measure the transmitted signal for the device $D_{1}$. An input
  microwave signal (RF) is sent to the sample through a highly
  attenuated microwave line. The output signal is first amplified by a
  cold amplifier and then by a series of room temperature amplifiers
  before being demodulated and recorded by an analog to digital
  converter (ADC). A bandpass filter (BP), a low pass filter (LP) and
  two circulators are used to prevent thermal and amplifier noise from
  reaching the sample.}
\label{fig:SI1}
\end{figure*}

Our devices consist of single-junction transmon qubits embedded in
aluminium 3D cavities thermally anchored at $10\,$mK in a dilution
refrigerator. We have measured two different devices, referred to as
$D_{1}$ and $D_{2}$, whose characteristic parameters can be found in
Table \ref{table:Tab1}. Fig. \ref{fig:SI1}(a) shows the device $D_1$
consisting of a two-ports Al microwave cavity which holds a
lithographically patterned Al transmon qubit [Fig. \ref{fig:SI1}(b)]
on a sapphire substrate. The Josephson junction was fabricated using a
double-angle shadow evaporation technique. Fig. \ref{fig:SI1}(c) shows
the experimental setup used for measuring $D_{1}$, a similar setup was
used for $D_{2}$. For the device $D_{1}$ $(D_{2})$ we measured the
transmitted (reflected) signal over many experiments using a
heterodyne detection scheme and computed the average signal.

Each cavity has a bare resonant frequency of $f_{c}=\omega_c/(2\pi)$
supporting a mode coupled to a transmon with the lowest transition
frequency $f_q \equiv f_{01}=\omega_{01}/(2\pi)$. The interaction
between the uncoupled qubit state $\Ket{n}$ and the cavity with
coupling strength $g$ causes a dispersive shift of the cavity
resonance $f_{c}$, depending on the state of the qubit.

\begin{table*}[h]
\begin{center}
\begin{tabular}{ |p{2cm}|p{3cm}|p{3cm}|p{2cm}|p{2cm}|p{2cm}|p{2cm}|}
\hline
 Device & $f_c$ (GHz) & $\delta/2\pi$ (GHz)&  $g/2\pi$ (GHz) & $E_J/E_C$ & $T_1$
 ($\mu$s) & $T_2$ ($\mu$s) \\ 
 \hline
 $D_1$  &   10.426  & 0.984 &  0.313 & 314 & 2.64 & 4.00\\
 $D_2$  &   10.567 & 2.383 &  0.335 & 165 & 2.20 & 2.10\\
 \hline
\end{tabular}
\caption{Characteristic parameters of the experimental devices: $f_c$
  is the bare cavity frequency, $\delta$ is the qubit-cavity detuning,
  $g$ is the coupling strength, $E_J/E_C$ is the ratio of the
  Josephson energy to the charging energy, $T_1$ is the relaxation and
  $T_2$ the dephasing time of the qubits.} {\label{table:Tab1}} 
\end{center}
\end{table*}

\begin{figure*}[!ht]
\includegraphics[width=6.5in]{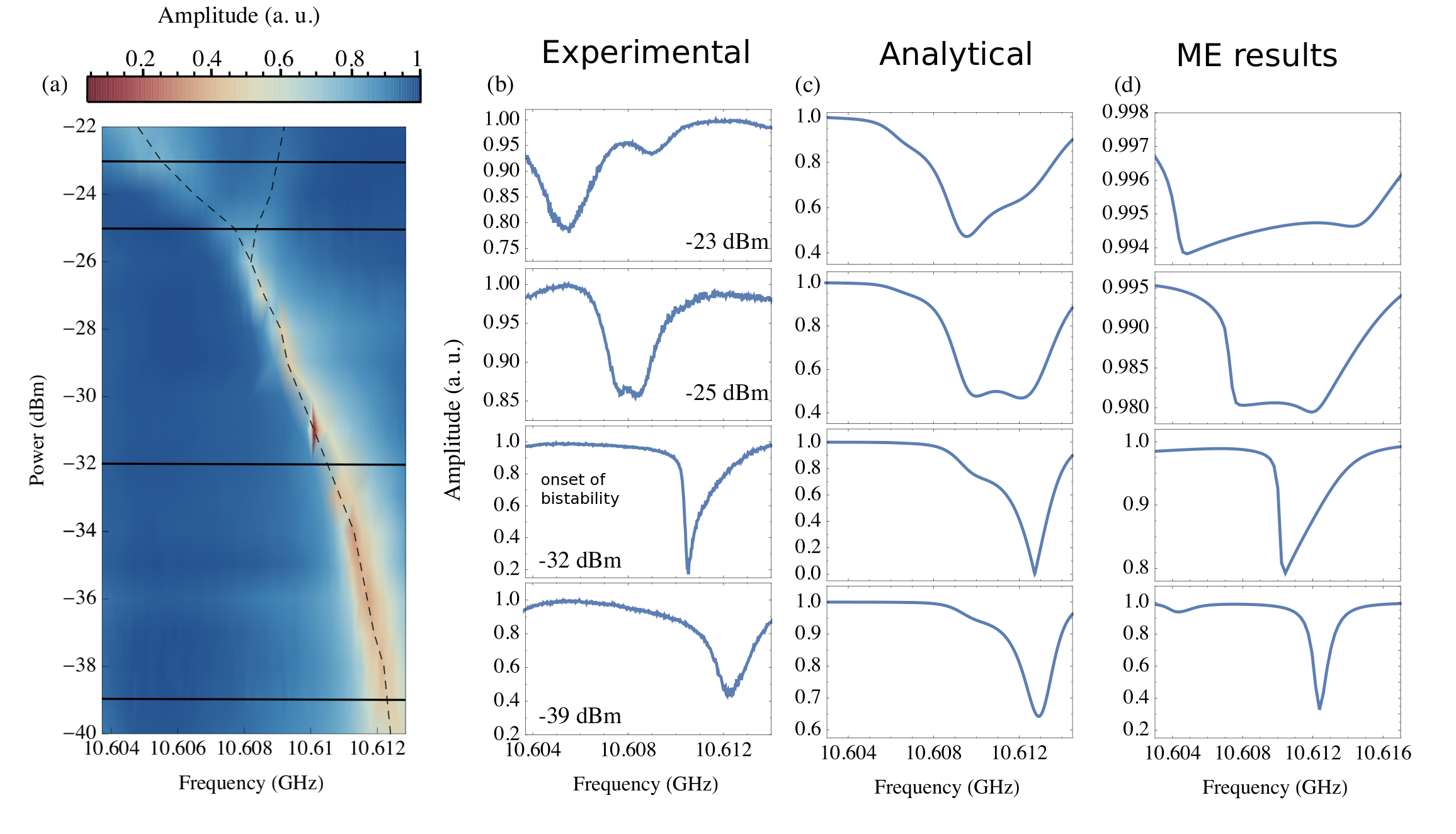}
\caption{Experimental and theoretical reflection curves for
    the ground state excitation of the device $D_2$, consisting of a
    3D cavity resonator coupled to a transmon qubit. The low power
  curves show a dispersive shift of $\chi^{(0)}_c=g^2/\delta$ from the
  bare cavity frequency. \textbf{(a)} Experimental reflection
  amplitude as a function of the drive frequency and power. At low
  driving power the response of the cavity is a Lorentzian centered at
  $10.612\,$GHz; as the power increases the Lorentzian minimum moves
  towards lower frequencies. Distortion of the Lorentzian shape occurs
  as the power is further increased as a result of bistability. At the
  power of $-32\,$dBm we mark the {\it onset of bistability}
  corresponding to the distorted Lorentzian shape with an abrupt
  wall-like profile. At a power of $-25\,$dBm we observe the emergence
  of two reflection features whose separation increases with
  increasing driving power. \textbf{(b)} Horizontal cuts through
  \textbf{(a)} at different drive powers ($-23\,$dBm, $-25\,$dBm,
  $-32\,$dBm, $-39\,$dBm). \textbf{(c)} Analytical results from the
  effective Fokker-Planck model. \textbf{(d)} ME simulations for a ten-level
  transmon with anharmonicity coefficient $\chi/2\pi=-242\,$MHz and
  dipole coupling strength $g/2\pi=335\,$ MHz.}
\label{fig:SI2}
\end{figure*}

In Fig. \ref{fig:SI2} we show the experimental reflection amplitude
for a different device ($D_2$, see Table \ref{table:Tab1} for more
details), as a function of the driving parameters, next to our
theoretical predictions. The experimental Lorentzian curve [Figs. \ref{fig:SI2}(a, b)] shifts towards lower frequencies with
increasing power, followed by the emergence of a dip separation, in
close agreement with the theoretical predictions shown in
Figs. \ref{fig:SI2}(c, d). The horizontal line at $-32\,$dBm marks the onset of
bistability, corresponding to a frequency shift of $\kappa$.
The analytical formulation already captures the pair of emerging dip
lines occurring for increasing drive strength, in close agreement with
the experiment. The ME results show more conspicuously the shift of
the resonance curve to the left for lower drive strengths, but lead to
a weaker separation because of thermal broadening.

\section{Theoretical Remarks}

\subsection{Formulation of the Master Equation}

The master equation (ME) corresponding to the open driven GJC model
consisting of a multilevel system (transmon qubit) coupled to a single
cavity mode with resonance frequency $\omega_c$ reads
\begin{equation}\label{ME}
\begin{aligned}
&\dot{\rho}=-i[H_{\textrm{GJC}}, \rho] + \kappa[\overline{n}(\omega_c)+1]
\mathcal{L}\{a, \rho\} + \kappa
\overline{n}(\omega_c)\mathcal{L}\{a^{\dagger}, \rho\} +\\ 
& + (\gamma/2)\mathcal{L}\left\{\sum_{j} \alpha_j \Ket{j} \Bra{j+1},
\rho\right\} +\\ 
&+(\gamma_{\phi}/2)\mathcal{L}\left\{\sum_j \beta_j \ket{j}\bra{j}, \rho\right\}, 
\end{aligned}
\end{equation}
where $\mathcal{L}\{A,\rho\}=2A\rho A^{\dagger} - A^{\dagger} A \rho -
\rho A^{\dagger} A$ is the Liouvillian super-operator acting upon the
Lindblad operator $A$, $\rho$ is the reduced density operator for the
cavity-atom system, $2 \kappa$ the cavity relaxation rate, $\gamma$
($\gamma_{\phi}$) the transmon qubit dissipation (dephasing) rate and
$\overline{n}(\omega_{c})=[e^{\omega_{c}/T}-1]^{-1}$ is the mean
photon number for an oscillator with frequency $\omega_{c}$ at
temperature $T$ ($k_B=1$) (we assume no thermal occupation for the
transmon qubit).  Expressions for the terms $\alpha_j$ and $\beta_j$
can be found in \cite{BishopRabi}.

The Hamiltonian $H_{\textrm{JC}}$ of the main text can also be
inserted as the system Hamiltonian in \eqref{ME}, where customarily
one uses pseudospin operators ($\sigma_{\pm}, \sigma_z$) to account
for the two-level system.  The dressed-cavity system Hamiltonian in
the Duffing approximation of the JC model, on the other hand, deserves
special attention because the generator of the decoupling
transformation required to decouple cavity and qubit does not commute
with the drive term and the super-operators transform non trivially.
As discussed in the main text, this approximation reads
\begin{equation}
\begin{aligned}
  &H_{\textrm{D}} = [\omega_c + g^4/\delta^3 - (g^2/\delta)\sigma_z +
  2 (g^4/\delta^3)\sigma_z + \\  
  &+ \mathcal{O}\left(g^6/\delta^5\right)]a^{\dagger}a +
  \left[g^4/\delta^3
    +\mathcal{O}\left(g^6/\delta^5\right)\right]\sigma_z{a^{\dagger}}^2
  a^2.
\end{aligned}
\end{equation}
The above operator may be inserted in \eqref{ME} for $\sigma_z=-1$
together with a drive term only to the leading order in $g/\delta$ and
$N^{-1/2}$ \cite{BishopJC}.
\begin{figure*}
\includegraphics[width=3.5in]{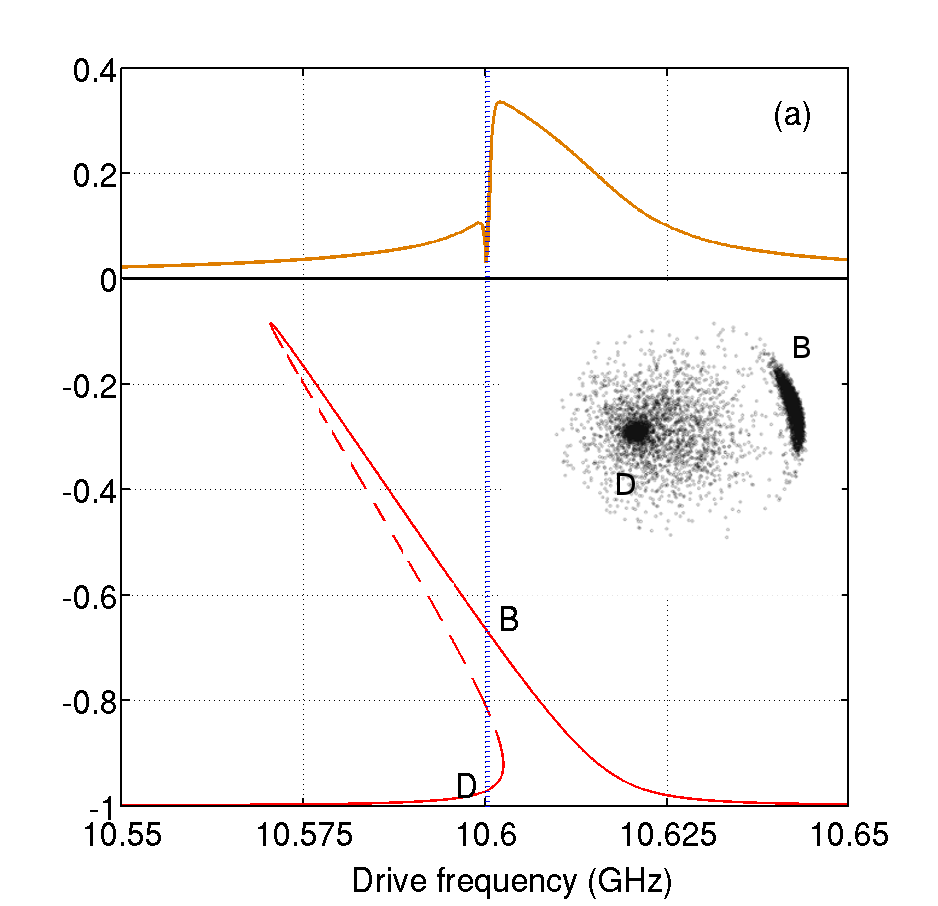}
\includegraphics[width=3.2in]{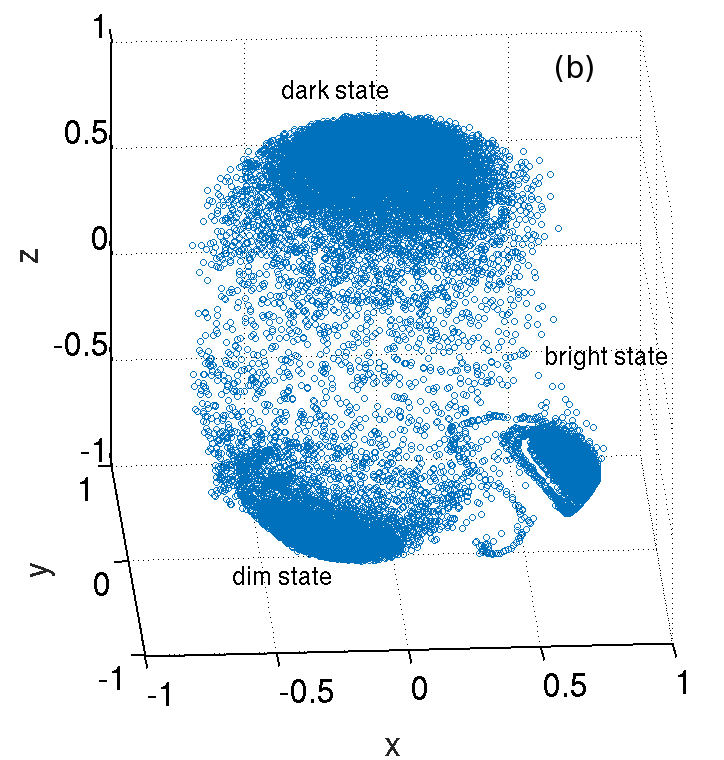}
\caption{Qubit response in the JC model: \textbf{(a)} The
  steady-state solution of the Maxwell-Bloch equations for
  $\braket{\sigma_z}$ (lower panel) together with the pseudospin
  projection $\left| \braket{\sigma_{-}} \right|$ (upper panel) from
  the full ME solution, showing the coherent cancellation dip. The
  inset shows the equatorial plane view of the qubit vector from the
  north pole of the Bloch sphere, with the two distinct distributions
  corresponding to the states B and D. The broken line indicates the
  driving frequency $\omega_d/2\pi=10.6005\,$GHz. \textbf{(b)} The
  accompanying scatter-plot of the average Bloch vector (for
  $\omega_d/2\pi=10.6005\,$GHz) at different times for a single
  quantum trajectory calculated from the reduced qubit density
  matrix.}
\label{fig:SI3} 
\end{figure*}

\subsection{Excitations in the JC oscillator}

\begin{figure}[!ht]
\includegraphics[width=3.4in]{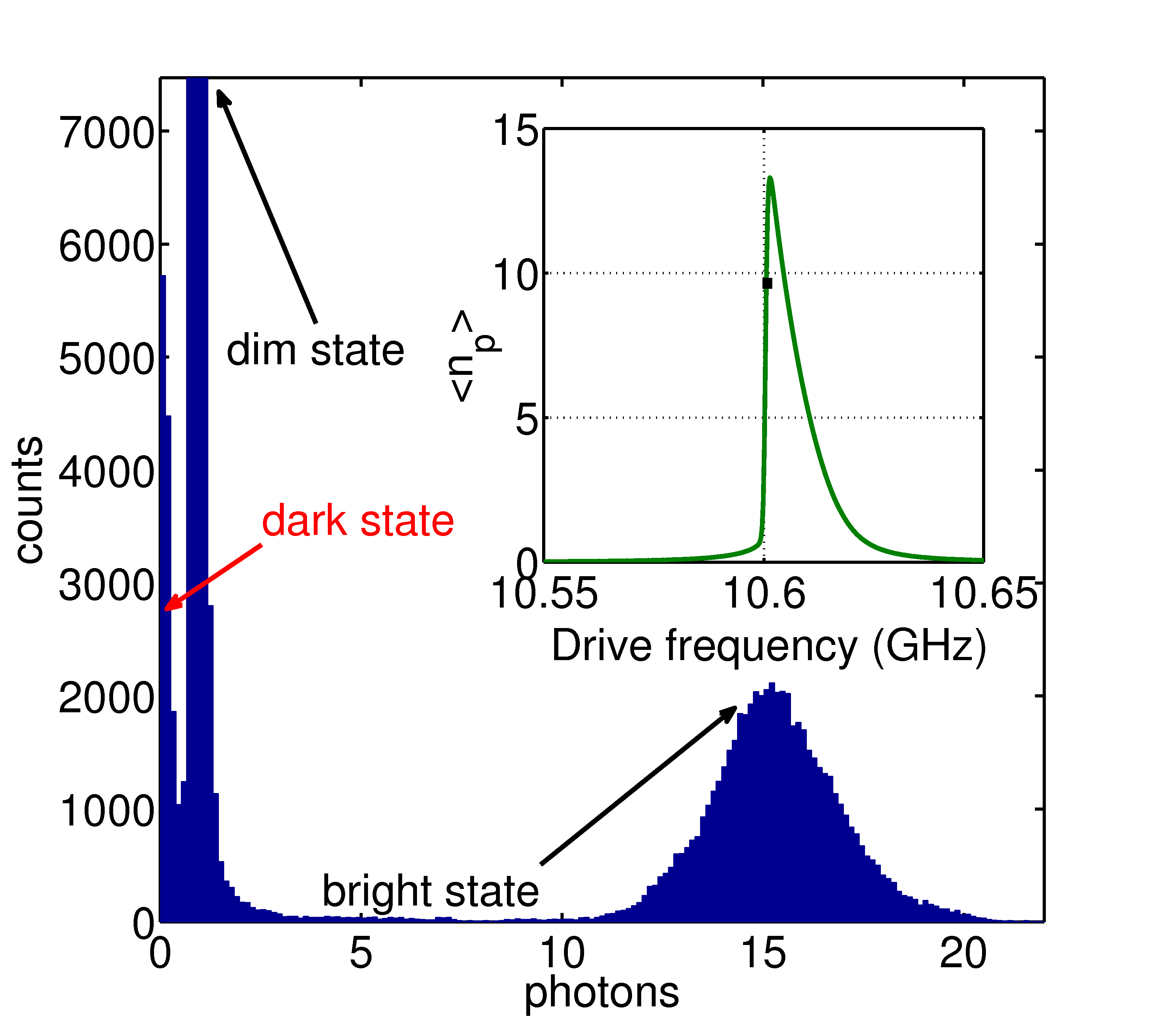}
\caption{Photon histogram corresponding to the quantum trajectory of Fig. \ref{fig:MEU}(b). The inset depicts $\braket{n_p}=\braket{a^{\dagger}a}$ as a function of the drive frequency for the same parameters used in Fig. \ref{fig:MEU}. The point on the curve marks the coordinates corresponding to the response depicted in Fig. \ref{fig:MEU}.}
\label{fig:SI4} 
\end{figure}

We will now focus on the behavior of the qubit during the bistable switching. In Fig \ref{fig:SI3}(a) we depict the qubit response in the driving regime where the Duffing approximation is not applicable. As mentioned in the main text, we can observe that $\left|\braket{\sigma_{-}}\right|$ exhibits the characteristic coherent cancellation dip. In Fig. \ref{fig:SI3}(b) we present the full scatter-plot for the qubit vector in the Bloch sphere, corresponding to the illustration of Fig. \ref{fig:MEU}(c) (but now including the data for the dark state). In the inset of Fig. \ref{fig:SI4} we show one of the two components (the cavity field) of the system total excitation, the eigenvalue of the operator $\mathcal{N}$ for the JC oscillator. None of the constituents of $\mathcal{N}$ ($n_p$ and $\sigma_{+}\sigma_{-}$) show the coherent cancellation dip that characterizes the corresponding complex amplitude. Instead, they both present a distorted Lorentzian profile, which marks the onset of amplitude bistability.

\begin{figure*}[!ht]
\centering
\includegraphics[width=3.5in]{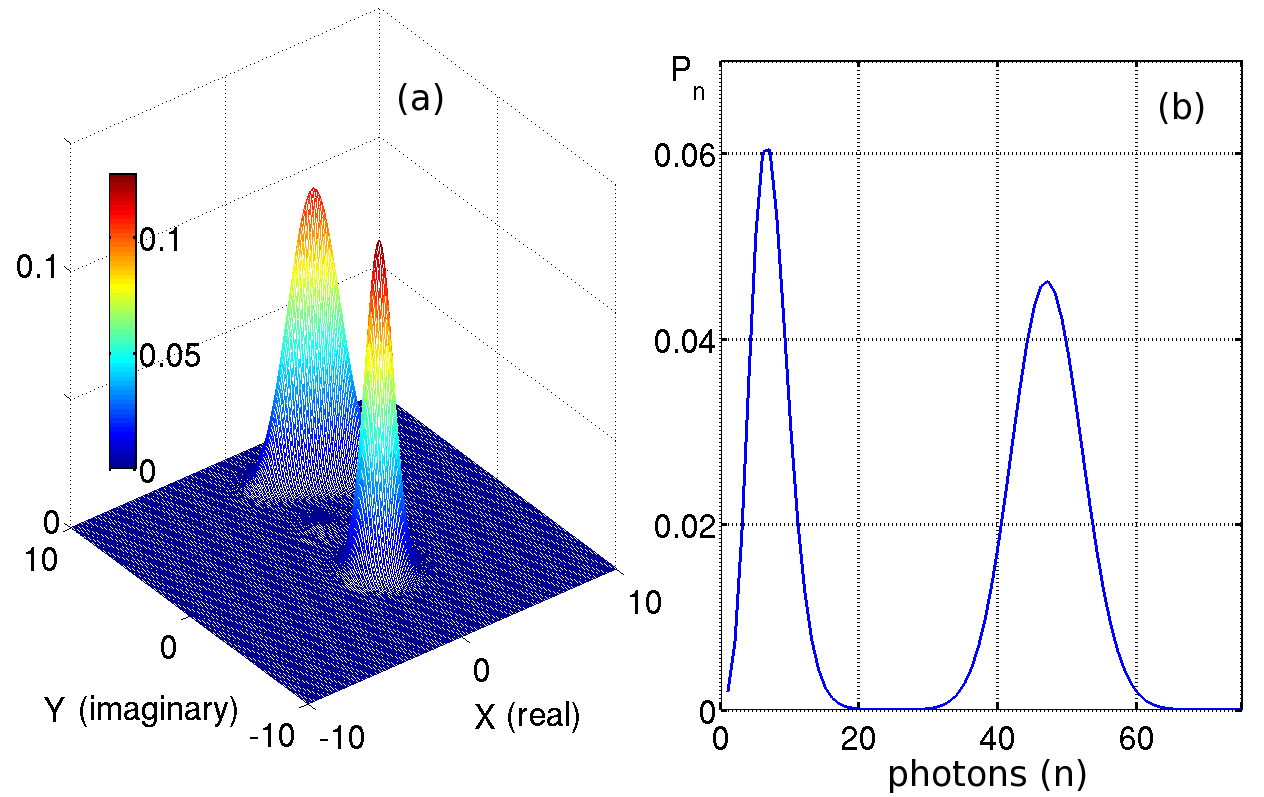}
\includegraphics[width=3.5in]{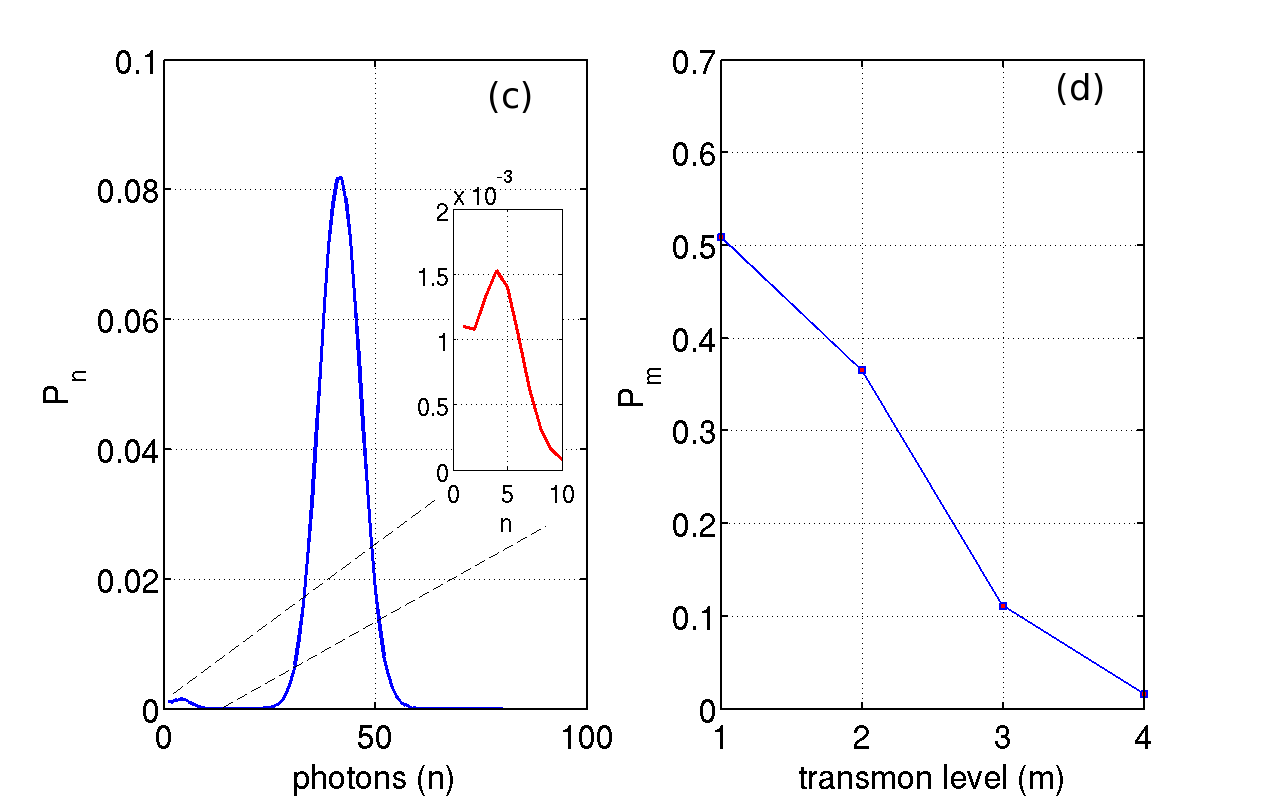}
\caption{ \textbf{(a)} $Q$ function and \textbf{(b)} $P_n$ distribution
   for a four-level transmon coupled to a cavity driven at a
  frequency $f_d=\omega_d/2\pi=10.6082\,$GHz prior to the coherent cancellation dip. \textbf{(c)} Probability distribution for the cavity occupancy, and
  \textbf{(d)} for the transmon, at the drive frequency $f_d=\omega_d/2\pi=10.6091\,$GHz past the position of the coherent cancellation dip.}
\label{fig:SI5}
\end{figure*}
\begin{figure}[!ht]
\includegraphics[width=3.3in]{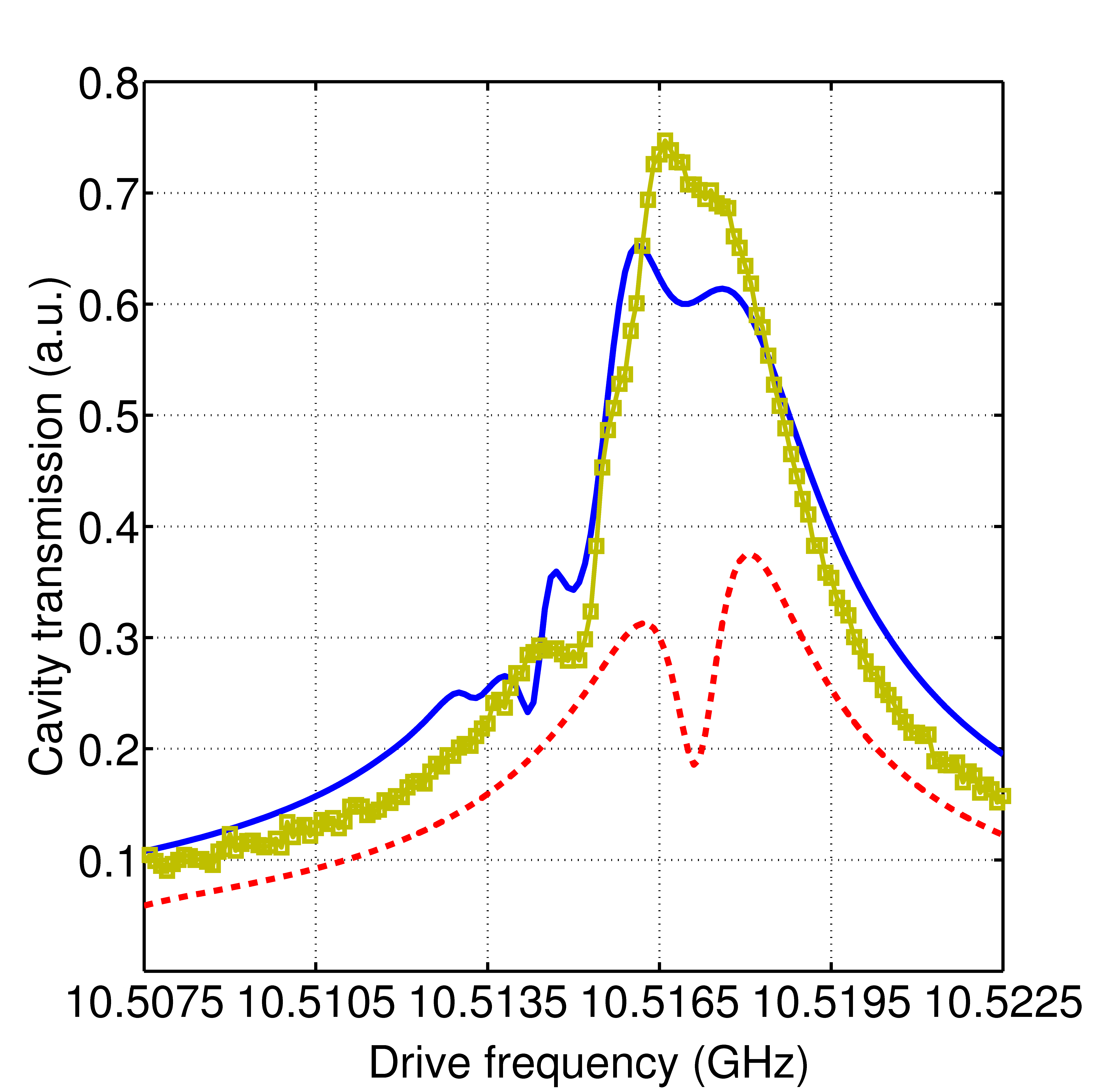}
\caption{Cavity transmission as a function of the drive frequency for the effective Fokker-Planck model. The experimental data for the driving power of $-46\,$dBm are compared to the predictions of Eq. \eqref{momentD} for $\chi/2\pi=-150\,$MHz (dashed red line) and $\chi/2\pi=-20\,$MHz (solid blue line).}
\label{fig:SI6} 
\end{figure}

\subsection{Bimodality in the GJC}

Bimodality is the underlying feature shaping the homodyne cavity
response in the driving regime we have considered in the main text for
the JC oscillator and its Duffing reduction. We show here that cavity
bimodality is also present for a 4-level transmon, as indicated by the
bimodal distribution in Fig. \ref{fig:SI5}(a). The emergence of bimodality can be also traced in the transmon level occupation depicted in Fig. \ref{fig:SI5}(b). The `dark' state is indicated by the departure from the `dim' state
Poissonian distribution at $P_0$ [inset of Fig. \ref{fig:SI5}(c)].

\subsection{Approaches to single-atom bistability}

Single-atom absorptive bistability was first reported at resonance in
\cite{SingleAtom}, where two distinct states were identified that are
long-lived in terms of the characteristic decay times of the two-level
atom and cavity. Adiabatic elimination of the atomic variables has been used in
\cite{Bonifacio} to extract approximate Gaussian distributions able to
adequately calculate mean square photon fluctuations for absorptive
optical bistability at resonance. However, adiabatic elimination underestimates the dynamical r\^{o}le of the qubit in shaping the system response.  There has been evidence
that the effect of the qubit on the cavity mode is significant in a
situation, where the drive is constantly tuned to maintain resonance
\cite{autoresonance}. The two degrees of freedom describing the system and the
inherent square root nonlinearity mark the departure from the Duffing
oscillator \cite{BishopJC}. Even in the dressed model describing
dynamic Rabi splitting for the driven JC oscillator
\cite{CarmichaelBook2}, with one degree of freedom, the effective
Hamiltonian leads to an equivalent description for the master equation
of resonance fluorescence which, as is well known, cannot be reduced
to a Fokker-Planck equation (FPE) \cite{CarmichaelBook1} following the
adiabatic elimination used in Ref. \cite{Bonifacio}.

\subsection{The effective Fokker-Planck model}

We will now provide more details for the adiabatic elimination of the
cavity in the transmon-cavity system. In this model we approximate the
level structure of the transmon itself by a Duffing oscillator. We set
$\omega_{n}=\omega_q n - \chi n(1-n)/2$ in Eq. \eqref{TransmonJC},
which is valid when the transmon occupation is relatively low. This
Duffing approximation is a consequence of retaining the lowest order
terms in the expansion of the Josephson potential
\cite{TransmonPaper}. We then adiabatically eliminate the cavity from
the system in the limit $2\kappa \gg \gamma$, and solve the
Fokker-Planck equation analytically for the steady-state distribution. The solution obtained retains
much of the structure of the full system despite the elimination
process, and also allows us to recover the moments of the cavity field
in the steady state. Thus, we model the transmon-cavity system as a
nonlinear oscillator, with the first moment given by the result in
Eq. \eqref{momentD}.

We can also use the intracavity amplitude to plot the reflection of
the cavity. In Fig. \ref{fig:SI2}(c) we show the calculated reflection
for the device $D_2$, compared with experimental results. For a
further comparison we solve numerically the corresponding ME for the
same parameters, and account for a finite temperature $T \simeq
200\,$mK in the presence of dissipation. For this device we see that
the features of the analytical reflection spectra agree qualitatively
with both experiment and ME at all four plotted powers.

For the device $D_1$, this analytical solution does not provide a good
fit to the experimental data, since we are no longer operating in the
limit $2\kappa \gg \gamma$. When $\kappa \approx \gamma$, this
solution appears to overestimate the effect of the transmon
nonlinearity on the cavity transmission spectrum. In
Fig. \ref{fig:SI6} we plot the cavity amplitude for different values
of the anharmonicity coefficient, comparing with the experimental
data. We see that for the true device anharmonicity of $\chi/2\pi=-150\,$MHz
the response resembles the JC oscillator in Fig. \ref{fig:GJCcomp}(b). If a much lower anharmonicity coefficient $\chi/2\pi=-20\,$MHz is selected, then
the response is in closer agreement with the GJC model and
experiment. This is consistent with the idea that the nonlinearity is
overestimated in this approximation -- if the experimental
nonlinearity happens to be large it is then additionally magnified by
the approximation so that the transmon behaves like a true two-level
system. The solution to make the model useful in this regime is to
choose an effective anharmonicity parameter $\chi$ smaller then the one
measured by the experiment, so that it is enhanced to the level of the true
experimental value by the approximation. 

\vspace{3mm}

\centerline{$\star \star \star \star \star$}


\begin{thebibliography}{43}
\expandafter\ifx\csname natexlab\endcsname\relax\def\natexlab#1{#1}\fi
\expandafter\ifx\csname bibnamefont\endcsname\relax
  \def\bibnamefont#1{#1}\fi
\expandafter\ifx\csname bibfnamefont\endcsname\relax
  \def\bibfnamefont#1{#1}\fi
\expandafter\ifx\csname citenamefont\endcsname\relax
  \def\citenamefont#1{#1}\fi
\expandafter\ifx\csname url\endcsname\relax
  \def\url#1{\texttt{#1}}\fi
\expandafter\ifx\csname urlprefix\endcsname\relax\def\urlprefix{URL }\fi
\providecommand{\bibinfo}[2]{#2}
\providecommand{\eprint}[2][]{\url{#2}}


\bibitem[{\citenamefont{Haroche}(2001)}]{cavQED}
\bibinfo{author}{\bibfnamefont{J.}~\bibnamefont{Raimond}},
\bibinfo{author}{\bibfnamefont{M.}~\bibnamefont{Brune}} and
\bibinfo{author}{\bibfnamefont{S.}~\bibnamefont{Haroche}},
  \bibinfo{journal}{Rev. Mod. Phys.} \textbf{\bibinfo{volume}{73}},
  \bibinfo{pages}{565} (\bibinfo{year}{2001}).

\bibitem[{\citenamefont{Koch}(2007)}]{TransmonPaper}
\bibinfo{author}{\bibfnamefont{J.}~\bibnamefont{Koch}} \textit{et al.},
\bibinfo{journal}{Phys. Rev. A} \textbf{\bibinfo{volume}{76}},
  \bibinfo{pages}{042319} (\bibinfo{year}{2007}).

\bibitem[{\citenamefont{Chiorescu}(2004)}]{cQED1}
\bibinfo{author}{\bibfnamefont{I.}~\bibnamefont{Chiorescu}} \textit{et al.},
  \bibinfo{journal}{Nature (London)} \textbf{\bibinfo{volume}{431}},
  \bibinfo{pages}{159} (\bibinfo{year}{2004}).

\bibitem[{\citenamefont{Wallraff}(2004)}]{cQED2}
\bibinfo{author}{\bibfnamefont{A.}~\bibnamefont{Wallraff}}, \textit{et al.},
  \bibinfo{journal}{Nature (London)} \textbf{\bibinfo{volume}{431}},
  \bibinfo{pages}{162} (\bibinfo{year}{2004}).

\bibitem[{\citenamefont{Nakamura}(1999)}]{JJ}
\bibinfo{author}{\bibfnamefont{Y.}~\bibnamefont{Nakamura}},
\bibinfo{author}{\bibfnamefont{Y.}~\bibfnamefont{A.}~\bibnamefont{Pashkin}} and
\bibinfo{author}{\bibfnamefont{J.}~\bibfnamefont{S.}~\bibnamefont{Tsai}}
  \bibinfo{journal}{Nature (London)} \textbf{\bibinfo{volume}{398}},
  \bibinfo{pages}{786} (\bibinfo{year}{1999}).
  
  \bibitem[{\citenamefont{WallsBook}(2010)}]{WallsBook}
\bibinfo{author}{\bibfnamefont{D.}~\bibnamefont{F.}~\bibnamefont{Walls}} and
\bibinfo{author}{\bibfnamefont{G.}~\bibnamefont{J.}~\bibnamefont{Milburn}},
  \bibinfo{book}\textit{Quantum Optics}~(\bibinfo{publisher}{Springer, Berlin}, \bibinfo{year}{2010}).

\bibitem[{\citenamefont{Aspelmeyer}(2014)}]{ReviewOptomechanics}
\bibinfo{author}{\bibfnamefont{M.}~\bibnamefont{Aspelmeyer}},
\bibinfo{author}{\bibfnamefont{T.}~\bibnamefont{J.}~\bibnamefont{Kippenberg}} and
\bibinfo{author}{\bibfnamefont{F.}~\bibnamefont{Marquardt}},
  \bibinfo{journal}{Rev. Mod. Phys.} \textbf{\bibinfo{volume}{86}},
  \bibinfo{pages}{1391} (\bibinfo{year}{2014}).

\bibitem[{\citenamefont{Carmichael}(2015)}]{PhotonBlockade}
\bibinfo{author}{\bibfnamefont{H.}~\bibnamefont{J.}~\bibnamefont{Carmichael}},
\bibinfo{journal}{Phys. Rev. X} \textbf{\bibinfo{volume}{5}},
  \bibinfo{pages}{031028} (\bibinfo{year}{2015}).


\bibitem[{\citenamefont{Rempe}(1991)}]{NatomsBist}
\bibinfo{author}{\bibfnamefont{G.}~\bibnamefont{Rempe}} \textit{et al.},
\bibinfo{journal}{Phys. Rev. Lett.} \textbf{\bibinfo{volume}{67}},
  \bibinfo{pages}{1727} (\bibinfo{year}{1991}).


\bibitem[{\citenamefont{Kerckhoff}(2011)}]{Remnants}
\bibinfo{author}{\bibfnamefont{J.}~\bibnamefont{Kerckhoff}},
\bibinfo{author}{\bibfnamefont{M.}~\bibfnamefont{A.}~\bibnamefont{Armen}} and
\bibinfo{author}{\bibfnamefont{H.}~\bibnamefont{Mabuchi}},
\bibinfo{journal}{Opt. Express} \textbf{\bibinfo{volume}{19}},
  \bibinfo{pages}{24468} (\bibinfo{year}{2011}).

\bibitem[{\citenamefont{Tian}(1992)}]{QuantumTrajectory}
\bibinfo{author}{\bibfnamefont{L.}~\bibnamefont{Tian}} and
\bibinfo{author}{\bibfnamefont{H.}~\bibnamefont{J.}~\bibnamefont{Carmichael}},
\bibinfo{journal}{Phys. Rev. A} \textbf{\bibinfo{volume}{46}},
  \bibinfo{pages}{R6801(R)} (\bibinfo{year}{1992}).

\bibitem[{\citenamefont{Bishop}(2009)}]{BishopRabi}
\bibinfo{author}{\bibfnamefont{L.}~\bibnamefont{S.}~\bibnamefont{Bishop}} \textit{et al.},
  \bibinfo{journal}{Nat. Phys.} \textbf{\bibinfo{volume}{5}},
  \bibinfo{pages}{105} (\bibinfo{year}{2009}).

\bibitem[{\citenamefont{JBA}(2009)}]{JBA}
\bibinfo{author}{\bibfnamefont{R.}~\bibnamefont{Vijay}},
\bibinfo{author}{\bibfnamefont{M.}~\bibfnamefont{H.}~\bibnamefont{Devoret}} and
\bibinfo{author}{\bibfnamefont{I.}~\bibnamefont{Siddiqi}},
  \bibinfo{journal}{Rev. Sci. Instrum.} \textbf{\bibinfo{volume}{80}},
  \bibinfo{pages}{111101} (\bibinfo{year}{2009}).


\bibitem[{\citenamefont{CarmichaelBook}(2008)}]{CarmichaelBook2}
\bibinfo{author}{\bibfnamefont{H.}~\bibnamefont{J.}~\bibnamefont{Carmnichael}},
  \bibinfo{book}\textit{Statistical Methods in Quantum Optics 2}~(\bibinfo{publisher}{Springer}, \bibinfo{year}{2008}).

\bibitem[{\citenamefont{Murch}(2012)}]{autoresonance}
\bibinfo{author}{\bibfnamefont{K.}~\bibnamefont{W.}~\bibnamefont{Murch}} \textit{et al.},
  \bibinfo{journal}{Phys. Rev. B} \textbf{\bibinfo{volume}{86}},
  \bibinfo{pages}{220503(R)} (\bibinfo{year}{2012}).


\bibitem[{\citenamefont{Boissonneault}(2010)}]{Boissonneault}
\bibinfo{author}{\bibfnamefont{M.}~\bibnamefont{Boissonneault}},
  \bibinfo{author}{\bibfnamefont{J.}~\bibnamefont{M.}~\bibnamefont{Gambetta}} and
  \bibinfo{author}{\bibfnamefont{A.}~\bibnamefont{Blais}},
  \bibinfo{journal}{Phys. Rev. Lett.} \textbf{\bibinfo{volume}{105}},
  \bibinfo{pages}{100504} (\bibinfo{year}{2010}).


\bibitem[{\citenamefont{Bishop}(2010)}]{BishopJC}
\bibinfo{author}{\bibfnamefont{L.}~\bibnamefont{S.}~\bibnamefont{Bishop}},
  \bibinfo{author}{\bibfnamefont{E.}~\bibnamefont{Ginossar}} and
  \bibinfo{author}{\bibfnamefont{S.}~\bibnamefont{M.}~\bibnamefont{Girvin}},
  \bibinfo{journal}{Phys. Rev. Lett.} \textbf{\bibinfo{volume}{105}},
  \bibinfo{pages}{100505} (\bibinfo{year}{2010}).

\bibitem[{\citenamefont{Reed}(2010)}]{Reed}
\bibinfo{author}{\bibfnamefont{M.}~\bibnamefont{D.}~\bibnamefont{Reed}} \textit{et al.},
  \bibinfo{journal}{Phys. Rev. Lett.} \textbf{\bibinfo{volume}{105}},
  \bibinfo{pages}{173601} (\bibinfo{year}{2010}).

\bibitem[{\citenamefont{DykmanSmelyanskii}(1988)}]{DykmanJETP}
\bibinfo{author}{\bibfnamefont{M.}~\bibnamefont{Dykman}} and
  \bibinfo{author}{\bibfnamefont{V.}~\bibnamefont{N.}~\bibnamefont{Smelyanskii}},
  \bibinfo{journal}{Sov. Phys. JETP} \textbf{\bibinfo{volume}{67}},
  \bibinfo{pages}{1769} (\bibinfo{year}{1988}).

\bibitem[{\citenamefont{DykmanBook}(2012)}]{DykmanBook}
\bibinfo{author}{\bibfnamefont{M.}~\bibnamefont{I.}~\bibnamefont{Dykman}},
  \bibinfo{book}\textit{Fluctuating Nonlinear Oscillators}~(\bibinfo{publisher}{Oxford University Press}, \bibinfo{year}{2012}).

\bibitem[{\citenamefont{Peano}(2010)}]{Peano}
\bibinfo{author}{\bibfnamefont{V.}~\bibnamefont{Peano}} and
  \bibinfo{author}{\bibfnamefont{M.}~\bibnamefont{Thorwart}},
  \bibinfo{journal}{Europhys. Lett.} \textbf{\bibinfo{volume}{89}},
  \bibinfo{pages}{17008} (\bibinfo{year}{2010}).


\bibitem[{\citenamefont{Kamenev}(2011)}]{Kamenevbook}
\bibinfo{author}{\bibfnamefont{A}~\bibnamefont{Kamenev}},
  \bibinfo{book}\textit{Field Theory of Non-equilibrium Systems}~(\bibinfo{publisher}{Cambridge University Press}, Cambridge, UK \bibinfo{year}{2011}).

\bibitem[{\citenamefont{Maier}(1992)}]{Maier}
\bibinfo{author}{\bibfnamefont{R.}~\bibnamefont{S.}~\bibnamefont{Maier}} and
  \bibinfo{author}{\bibfnamefont{D.}~\bibnamefont{L.}~\bibnamefont{Stein}},
  \bibinfo{journal}{Phys. Rev. Lett.} \textbf{\bibinfo{volume}{69}},
  \bibinfo{pages}{3691} (\bibinfo{year}{1992}).

\bibitem[{\citenamefont{Dykman}(1994)}]{DykmanPLA}
\bibinfo{author}{\bibfnamefont{M.}~\bibnamefont{I.}~\bibnamefont{Dykman}},
  \bibinfo{author}{\bibfnamefont{M.}~\bibnamefont{M.}~\bibnamefont{Millonas}} and
  \bibinfo{author}{\bibfnamefont{V.}~\bibnamefont{N.}~\bibnamefont{Smelyanskiy}},
  \bibinfo{journal}{Phys. Lett. A} \textbf{\bibinfo{volume}{195}},
  \bibinfo{pages}{53} (\bibinfo{year}{1994}).

\bibitem[{\citenamefont{Graham}(1984)}]{Graham}
\bibinfo{author}{\bibfnamefont{R.}~\bibnamefont{Graham}} and
  \bibinfo{author}{\bibfnamefont{T.}~\bibnamefont{T\'{e}l}},
  \bibinfo{journal}{Phys. Rev. Lett.} \textbf{\bibinfo{volume}{52}},
  \bibinfo{pages}{9} (\bibinfo{year}{1984}).

\bibitem[{\citenamefont{Paik}(2011)}]{Paik}
\bibinfo{author}{\bibfnamefont{H.}~\bibnamefont{Paik}} \textit{et al.},
\bibinfo{journal}{Phys. Rev. Lett.} \textbf{\bibinfo{volume}{107}},
  \bibinfo{pages}{240501} (\bibinfo{year}{2011}).

\bibitem[{\citenamefont{Drummond}(1980)}]{DrummondWallsKerr}
\bibinfo{author}{\bibfnamefont{P.}~\bibnamefont{D.}~\bibnamefont{Drummond}} and
\bibinfo{author}{\bibfnamefont{D.}~\bibnamefont{F.}~\bibnamefont{Walls}},
  \bibinfo{journal}{J. Phys. A} \textbf{\bibinfo{volume}{13}},
  \bibinfo{pages}{725} (\bibinfo{year}{1980}).

\bibitem{doublerate} Here we denote the photon loss rate by $2\kappa$, in alignment with the rate equations of laser theory.

\bibitem[{\citenamefont{Carbonaro}(1979)}]{DecouplingT}
\bibinfo{author}{\bibfnamefont{P.}~\bibnamefont{Carbonaro}},
\bibinfo{author}{\bibfnamefont{G.}~\bibnamefont{Compagno}} and
\bibinfo{author}{\bibfnamefont{F.}~\bibnamefont{Persico}},
  \bibinfo{journal}{Phys. Lett.} \textbf{\bibinfo{volume}{73A}},
  \bibinfo{pages}{97} (\bibinfo{year}{1979}).

\bibitem{ComparisonRD} This expansion brings us to the fundamental $\sqrt{N}$
  nonlinearity of the driven JC oscillator \cite{SqrtN} with $N$
  excitations.  The oscillator response has been studied more
  extensively at resonance, where single-atom absorptive optical
  bistability was first reported in \cite{SingleAtom}. The $\sqrt{N}$
  anharmonic oscillator is invoked \cite{AnharmonicOscillator}
  alongside two excitation ladders, each associated with a bosonic
  field \cite{CarmichaelBook2}. Strong driving leads to spontaneous
  symmetry breaking and phase bistability following a second-order
  quantum phase transition \cite{AlsingCarmichael}. In the dispersive
  regime, on the other hand, the $\sqrt{N}$ eigenvalue dependence
  allows for a perturbation expansion in powers of
  $N/N_{\textrm{crit}}$, where $N_{\textrm{crit}}=\delta^2/(4g^2)$ is
  a characteristic system parameter, indicating the driving region
  where nonlinearity is important. The expansion is carried out
  around a critical point in the mean-field phase diagram
  \cite{BishopJC}.

\bibitem[{\citenamefont{Fink}(2008)}]{SqrtN}
\bibinfo{author}{\bibfnamefont{J.}~\bibnamefont{Fink}} \textit{et al.},
\bibinfo{journal}{Nature (London)} \textbf{\bibinfo{volume}{454}},
  \bibinfo{pages}{315} (\bibinfo{year}{2008}).

\bibitem[{\citenamefont{Savage}(1988)}]{SingleAtom}
\bibinfo{author}{\bibfnamefont{C.}~\bibnamefont{M.}~\bibnamefont{Savage}} and
\bibinfo{author}{\bibfnamefont{H.}~\bibnamefont{J.}~\bibnamefont{Carmichael}},
\bibinfo{journal}{IEEE J. Quantum Electron.} \textbf{\bibinfo{volume}{24}},
  \bibinfo{pages}{1495} (\bibinfo{year}{1988}).

\bibitem[{\citenamefont{Chough}(1996)}]{AnharmonicOscillator}
\bibinfo{author}{\bibfnamefont{Y.}~\bibnamefont{T.}~\bibnamefont{Chough}} and
\bibinfo{author}{\bibfnamefont{H.}~\bibnamefont{J.}~\bibnamefont{Carmichael}},
\bibinfo{journal}{Phys. Rev. A} \textbf{\bibinfo{volume}{54}},
  \bibinfo{pages}{1709} (\bibinfo{year}{1996}).


\bibitem[{\citenamefont{Alsing}(1991)}]{AlsingCarmichael}
\bibinfo{author}{\bibfnamefont{P.}~\bibnamefont{Alsing}} and
  \bibinfo{author}{\bibfnamefont{H.}~\bibnamefont{J.}~\bibnamefont{Carmichael}},
  \bibinfo{journal}{Quantum Opt.} \textbf{\bibinfo{volume}{3}},
  \bibinfo{pages}{13} (\bibinfo{year}{1991}).


\bibitem{driveterms}The drive terms can also be shown to remain
  linear in the cavity operators for low enough drives
  \cite{BishopJC}.

\bibitem[{\citenamefont{Blais}(2004)}]{Blais2004}
\bibinfo{author}{\bibfnamefont{A.}~\bibnamefont{Blais}} \textit{et al.},
\bibinfo{journal}{Phys. Rev. A} \textbf{\bibinfo{volume}{69}},
  \bibinfo{pages}{062320} (\bibinfo{year}{2004}).

\bibitem[{\citenamefont{BreuerBook}(2002)}]{BreuerBook}
\bibinfo{author}{\bibfnamefont{H.-P.}~\bibnamefont{Breuer}} and
\bibinfo{author}{\bibfnamefont{F.}~\bibnamefont{Petruccione}},
  \bibinfo{book}\textit{The Theory of Open Quantum Systems}~(\bibinfo{publisher}{Oxford University Press}, \bibinfo{year}{2002}).

\bibitem[{\citenamefont{PlatenBook}(1995)}]{PlatenBook}
\bibinfo{author}{\bibfnamefont{P.}~\bibnamefont{E.}~\bibnamefont{Kloeden}} and
\bibinfo{author}{\bibfnamefont{E.}~\bibnamefont{Platen}},
  \bibinfo{book}\textit{Stochastic Modelling and Applied Probability}~(\bibinfo{publisher}{Springer}, \bibinfo{year}{1995}).

\bibitem[{\citenamefont{Protocol}(2010)}]{Protocol}
\bibinfo{author}{\bibfnamefont{E.}~\bibnamefont{Ginossar}},
\bibinfo{author}{\bibfnamefont{L.}~\bibnamefont{S.}~\bibnamefont{Bishop}},
\bibinfo{author}{\bibfnamefont{D.}~\bibnamefont{I.}~\bibnamefont{Schuster}}, and
\bibinfo{author}{\bibfnamefont{S.}~\bibnamefont{M.}~\bibnamefont{Girvin}},
\bibinfo{journal}{Phys. Rev. A} \textbf{\bibinfo{volume}{82}},
  \bibinfo{pages}{022335} (\bibinfo{year}{2010}).
  
  \bibitem[{\citenamefont{Zhirov}(2008)}]{Zhirov}
\bibinfo{author}{\bibfnamefont{O.}~\bibnamefont{V.}~\bibnamefont{Zhirov}}
and
\bibinfo{author}{\bibfnamefont{D.}~\bibnamefont{L.}~\bibnamefont{Shepelyansky}},
  \bibinfo{journal}{Phys. Rev. Lett.} \textbf{\bibinfo{volume}{100}},
  \bibinfo{pages}{014101} (\bibinfo{year}{2008}).


\bibitem[{\citenamefont{Drummond}(1981)}]{DrummondWallsHG}
\bibinfo{author}{\bibfnamefont{P.}~\bibnamefont{D.}~\bibnamefont{Drummond}},
\bibinfo{author}{\bibfnamefont{K.}~\bibnamefont{J.}~\bibnamefont{McNeil}} and
\bibinfo{author}{\bibfnamefont{D.}~\bibnamefont{F.}~\bibnamefont{Walls}},
  \bibinfo{journal}{Opt. Acta} \textbf{\bibinfo{volume}{28}},
  \bibinfo{pages}{211} (\bibinfo{year}{1981}).

\bibitem[{\citenamefont{Elliott}(2016)}]{ElliottElimination}
\bibinfo{author}{\bibfnamefont{M.}~\bibnamefont{Elliott}} and
\bibinfo{author}{\bibfnamefont{E.}~\bibnamefont{Ginossar}},
\bibinfo{journal}{Phys. Rev. A} \textbf{\bibinfo{volume}{94}},
  \bibinfo{pages}{043840} (\bibinfo{year}{2016}).
  
\bibitem{SurreyDep} DOI: \url{https://doi.org/10.15126/surreydata.00813140}.

\bibitem{Bonifacio}
R. Bonifacio, M. Gronchi and L. A. Lugiato, Phys. Rev. A, \textbf{18}, 2266 (1978).

\bibitem{CarmichaelBook1}
H. J. Carmichael, \textit{Statistical Methods in Quantum Optics 1}, (Springer, New York, 1999).

 

\end{thebibliography}
\end{document}